\documentclass[aps,pra,twocolumn,superscriptaddress,showpacs,10pt]{revtex4-2}
\usepackage{amsmath}
\usepackage[dvips,xdvi]{graphicx}
\usepackage{amssymb}
\usepackage{epsfig}
\usepackage{xcolor}
\usepackage{physics}
\usepackage{soul,color}
\usepackage{microtype}

\begin{document}
	\title{Detection of Spin-Spatial-Coupling-Induced Dynamical Phase Transitions in Real Time}
	\author{J. O. Austin-Harris}
	\email{jared.o.austin@okstate.edu}
	\author{Z. N. Hardesty-Shaw}
    \author{C. Binegar}
    \author{P. Sigdel}
    \author{T. Bilitewski}
    \author{Y. Liu}
	\affiliation{Department of Physics, Oklahoma State University, Stillwater, Oklahoma 74078, USA}
	\date{\today}
    
	\begin{abstract}
    We demonstrate the real-time detection of dynamical phase transitions (DPTs) in lattice-confined spinor gases subject to \textit{a priori} unknown time-variant interactions, via the temporal behaviors of both the system energy and spinor phases extracted from the observed spin dynamics. Using this technique, we describe the observed nonequilibrium spin dynamics, governed by intricate spin-spatial couplings, across a range of conditions. This work also introduces an observable that can quickly identify DPTs at holding times when commonly-used order parameters still exhibit transient, nonuniversal behavior. Our approach can naturally extend to other complex systems subject to time-dependent parameters, such as Floquet systems under driven magnetic fields, driven interactions, or spin-flopping fields, with potential applications in the study of DPTs in nonintegrable models.
    \end{abstract}
	\maketitle
    \section{Introduction}
Ultracold spinor gases, highly-controllable quantum systems with a spin degree of freedom, have been proposed as ideal platforms for studying nonequilibrium phenomena~\cite{Stamper2013,Ueda2012,Zhang2005,Chang2005,Austin3,Zach1,Chen2019,Austin1,Austin2,Qingze2025,Yingmei2019,Guan2021,Marino2022,Zhou2023,Feldmann2021}. When out of equilibrium, spinor gases display spin oscillations driven by the competition of the quadratic Zeeman energy $q$ and the spin-dependent interactions $c_2$~\cite{Stamper2013,Ueda2012,Chen2019,Austin1,Austin2,Zach1,Zhang2005,Chapman2005PRL,Chang2005,Black2007,Yingmei2009,Yingmei2009_2,Pechkis2013,Lichao2014,Lichao2015,Austin3,Austin5,Kronjager2006,Jiang2014,Austin4}.  These spin oscillations reveal a rich dynamical phase diagram hosting dynamical phase transitions (DPTs)~\cite{Yingmei2019,Guan2021,Marino2022,Zhou2023,Feldmann2021,Robert2021,Dag2018}, that can be engineered through various modifications to system parameters.

DPTs are of fundamental interest due to their association with universal non-equilibrium critical phenomena and as pathways to quantum-enhanced sensing and quantum entanglement~\cite{Yingmei2019,Guan2021,Marino2022,Zhou2023,Feldmann2021,Robert2021,Dag2018}. There are two categories of DPTs recognized in the literature: type I DPTs, which display nonanalytic behavior in the steady state of a local order parameter, and type II DPTs, which display nonanalytic temporal behavior in a global order parameter after a quench~\cite{Yingmei2019,Guan2021,Marino2022,Zhou2023}. The study of DPTs has been accomplished both experimentally and theoretically in a wide variety of systems, including condensed matter systems~\cite{Marino2022,Shimano2020,Matsunaga2013}, trapped ions~\cite{Flaschner2018,Jurcevic2017,Zhang2017,Guan2021,Robert2021,Marino2022}, and both scalar and spinor ultracold gases~\cite{Dag2018,Yingmei2019,Qingze2025,Guan2021,Marino2022,Zhou2023,Feldmann2021,Austin5}. However, these investigations have thus far been primarily limited to systems that are well understood theoretically, such as integrable systems~\cite{Zhang2017,Robert2021,Marino2022}.

In this work, we demonstrate real-time detection of type II DPTs first in a simple well-understood system, i.e., spinor gases in free space, and then in a lattice-confined spinor system with \textit{a priori} unknown time-variant interactions, via temporal behaviors of spinor phases and system energy extracted from observed spin dynamics.  Realizing DPT detection in systems with unknown time-dependent system parameters opens an avenue for the study of crossover phenomena, universality, and DPTs in nonintegrable models, such as in spinor gases subject to weak spin-flopping fields, which can host quantum scars and quantum many-body scars~\cite{Marino2022,Robert2021,Austin4,Austin5,Dag2024}. %
We also introduce a new observable, the cutoff time $t_c$, that can quickly identify DPTs as, unlike commonly-used order parameters, it does not require multiple experimental runs and observations over at least one full period of the spin dynamics. The success of our informed predictions, especially in explaining the observed spin dynamics manipulated by intricate spatial dynamics over a range of conditions, suggests that similar techniques can be applied to analyze observations in other complex systems with time-dependent parameters, e.g., Floquet systems under driven quadratic Zeeman shift $q$~\cite{Fujimoto2019,Evrard2019,Li2019,Liu2022}, driven interaction $c_2$~\cite{Austin3, Qingze2025}, or resonant spin-flopping fields~\cite{Dag2024,Austin4,Austin5}.

\section{Experimental Sequence}
\begin{figure}[b]
    \includegraphics[width=86mm]{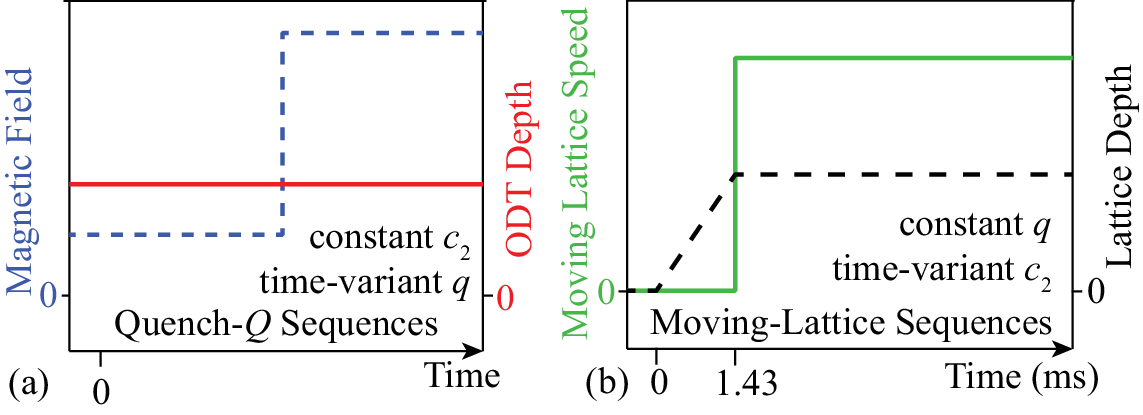}
    \caption{(a) Schematic of Quench-\textit{Q} sequences showing how the magnetic field (blue) and ODT trapping depth (red) vary in time. (b) Schematic of moving-lattice sequences showing how the lattice speed (green) and lattice depth (black) vary in time. Axes are not to scale in both panels.}\label{ExpSeq}
\end{figure}
\begin{figure*}[t]
	\includegraphics[width=176mm]{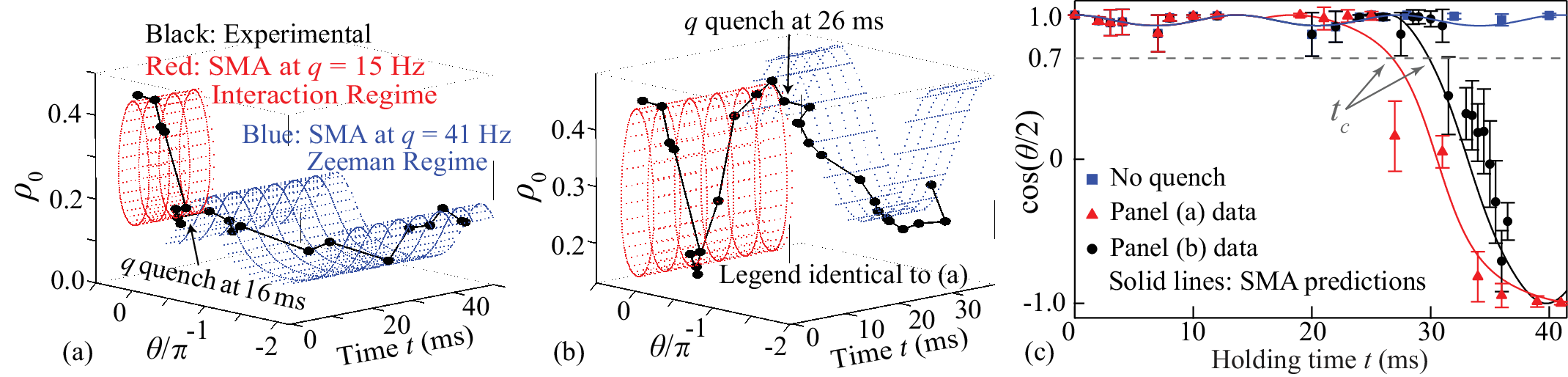}
    \caption{(a) Black circles show a DPT from an interaction regime to a Zeeman regime observed after a Quench-\textit{Q} sequence, a sudden quench in $q$ from 15~Hz to 41~Hz at $16~\mathrm{ms}$ with the timing of the quench roughly corresponding to a minimum in the $\rho_0$ oscillation where $\rho_0\approx0.12$ and $\theta\sim0$. Red (blue) dotted surfaces display the SMA equal energy contours of the interaction (Zeeman) regime occurring before (after) the $q$ quench for our system where $c_2=22$~Hz and $M=0$ (see Eq.~\eqref{MF_Ham}). (b) Similar to panel (a) but taken with the quench at $26~\mathrm{ms}$ roughly corresponding to a maximum in the $\rho_0$ oscillation where $\rho_0\approx0.42$ and $\theta\sim0$. (c) Triangles (circles) show that the extracted $\cos(\theta/2)$ identifies the DPTs in real time after the $q$ quenches at $16~\mathrm{ms}$ ($26~\mathrm{ms}$), while squares indicate no DPT occurs without a $q$ quench. Solid lines are SMA predictions, while the intersection of the dashed and solid lines marks the cutoff time $t_c$ where the system is conclusively shown to be in the Zeeman regime (see text).} \label{qQuench}
\end{figure*} 
Each experimental cycle begins with an $F=1$ spinor BEC of up to $10^5$ sodium atoms in a crossed optical dipole trap (ODT).  At holding time $t=0$ we prepare an initial state with $\rho_0(0)\approx0.45$, $M(0)=0$, and $\theta(0)=0$. Here  $\theta=\theta_1+\theta_{-1}-2\theta_0$ is the relative spinor phase, $\theta_{m_F}$ ($\rho_{m_F}$) is the phase (fractional population) of the $m_F$ hyperfine spin state. We note that the magnetization $M=\rho_{1} - \rho_{-1}$ is conserved when no external driving field is applied. %
We detect DPTs in real time induced via one of two experimental sequences, Quench-\textit{Q} and moving-lattice sequences  (see Fig.~\ref{ExpSeq}). These sequences result in time-variant ratios $c_2/q$ enabling us to engineer the dynamical phase diagram. Quench-\textit{Q} sequences use magnetic field quenches to induce a time-variant $q$ as a control parameter. Moving-lattice sequences impart an \textit{a priori} unknown time variance to $c_2$ using a moving lattice constructed from two nearly orthogonal lattice beams that are skew to the ODT beams. By quenching the frequency difference between the lattice beams from zero to $\Delta f$, which accelerates the lattice speed to $v=\lambda_{L} \Delta f$, the moving lattice near resonantly couples the $\mathbf{p}=0$ and $\mathbf{p}=2\hbar \mathbf{k}_L$ momentum states and induces a time-variant $c_2$ via the coupling of spin and spatial degrees of freedom~\cite{Zach1,Zach2}.  Here $\lambda_{L}/2\approx810~\mathrm{nm}$ is the lattice spacing, $\mathbf{k}_L$ is the lattice wave vector, and $h$ ($\hbar$) is the (reduced) Planck constant. At the end of an experimental cycle, atoms are released from all trapping potentials for ballistic expansion and spin-resolved imaging.

\section{Model}
For the data presented in this work, all spin states appear to share a common but potentially time-dependent spatial mode. Combined with the calculated spin healing length ($\approx12~\mu$m) being larger than the Thomas-Fermi radii ($\approx(9,9,7)~\mu$m) for all systems studied in this work, this supports use of a dynamical single spatial-mode approximation (SMA) to express the system Hamiltonian~\cite{Zach1}:
	\begin{equation}
		H/h\!=\!c_2\rho_0[1-\rho_0\!+\!\sqrt{(1\!-\!\rho_0)^2\!-\!M^2}\cos(\theta)]\!+\!q(1-\rho_0).\label{MF_Ham}
	\end{equation} 
 With the exception that $c_2$ and $q$ may be time dependent rather than strictly constant, Eq.~\eqref{MF_Ham} is identical to the well-known SMA-based Hamiltonian of $F=1$ spinor gases in free space~\cite{Zhang2005,Chang2005,Lichao2014,Austin5, Stamper2013,Pechkis2013,Kronjager2006,Black2007,Yingmei2009,Yingmei2009_2,Jiang2014,Lichao2015,Zach1,Ueda2012, Austin3}. %
 %Although the SMA was considered to only be valid when the spatial mode is frozen (i.e., $c_2$ is constant), recent experiments and theoretical work have demonstrated that this restriction is not strictly necessary and the SMA remains applicable in the more general case that all spin states share the same time-dependent spatial mode~\cite{Zach1,Austin3,Qingze2025}.
 Although the SMA was traditionally considered valid only for frozen spatial modes, recent work demonstrates that it remains applicable if all spin states share the same time-dependent spatial mode~\cite{Zach1,Austin3,Qingze2025}. %
 This is the case for our moving-lattice system, for which the observed complex spatial dynamics are nearly identical for all spin states. 
 
 Eq.~\eqref{MF_Ham} results in the following equations of motion for $\rho_0$ and $\theta$ \cite{Zhang2005,Chang2005,Lichao2014,Black2007,Austin5},
    \begin{align}
       \pdv{\rho_0}{t}=\frac{-2}{\hbar}\pdv{H}{\theta}=&\frac{c_2}{\pi}\rho_0\sqrt{(1-\rho_0)^2-M^2}\sin(\theta)\label{RhoEqOfMotion}\\
       \pdv{\theta}{t}=\phantom{-}\frac{2}{\hbar}\pdv{H}{\rho_0}=&\frac{c_2}{\pi}\frac{(1-\rho_0)(1-2\rho_0)-M^2}{\sqrt{(1-\rho_0)^2-M^2}}\cos(\theta)\notag\\ &+\frac{c_2}{\pi}(1-2\rho_0)-\frac{q}{\pi}.\label{ThetaEqOfMotion}
    \end{align}  
By approximating time derivatives as discrete differences evaluated based on the observed population dynamics, the equations of motion can be consistently solved for $c_2$ and $\theta$ to construct a complete picture of the full quantum dynamics (see Ref.~\cite{Austin5}). Two distinct dynamical regimes are predicted by these equations of motion: an interaction-dominated
(Zeeman-dominated) regime where $\theta$ is bounded (unbounded), as shown in Fig.~\ref{qQuench}~\cite{Austin3, Zach1}.

\begin{figure}[tb]
    \includegraphics[width=86mm]{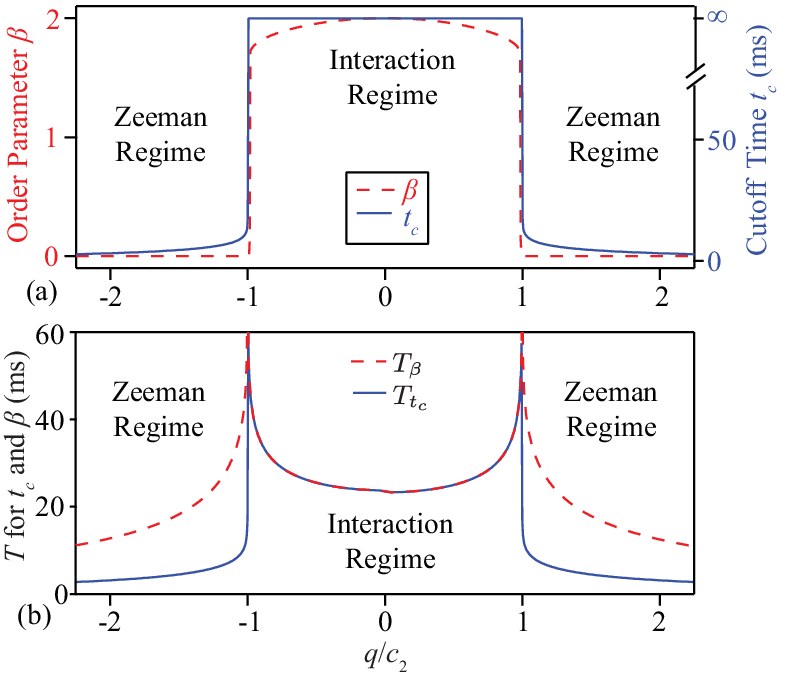}
    \caption{(a) The similarities in the predicted dependence of the cutoff time $t_c$ (solid line) and order parameter $\beta$ (dashed line) on $q/c_2$ confirm that $t_c$ is a good observable for detecting DPTs. (b) $T_{t_c}$ (solid line), the time needed to determine the cutoff time $t_c$, is always shorter than or equal to $T_{\beta}$ (dashed line), the time needed to determine $\beta$. In both panels, predictions are based on Eq.~\eqref{MF_Ham} for our system using a typical initial state in which $\rho_0(0)=0.5$ and $\theta(0)=0$. } \label{tcVsBeta}
\end{figure}  
\section{Real-time detection of DPTs}
Our experimental data in Fig.~\ref{qQuench}(a) and Fig.~\ref{qQuench}(b) reveal DPTs induced by quenching $q$ at approximately the minimum ($\rho_0\approx0.12$) and maximum ($\rho_0\approx0.42$)) of the pre-quench $\rho_0$ oscillation, respectively. Initially, the system is in the interaction regime with bounded $\theta$ (following a closed path in the phase diagram). After the $q$ quench, $\theta$ becomes unbounded marking a DPT to the Zeeman regime. These observations can be described by Eq.~\eqref{MF_Ham} with a constant $c_2\approx22~\mathrm{Hz}$ and $q=15~\mathrm{Hz}$ before ($q=41~\mathrm{Hz}$ after) the quench, as shown by the red (blue) energy contours in Figs.~\ref{qQuench}(a) and \ref{qQuench}(b). Here $\theta$ is extracted from the observed spin population dynamics using Eq.~\eqref{RhoEqOfMotion} and Eq.~\eqref{ThetaEqOfMotion} by approximating $\pdv{\rho_0}{t}$ as the discrete difference between $\rho_0$ data points and minimizing the difference between $\pdv{\theta}{t}$ and the discrete difference between $\theta$ points. This extends the technique developed in our prior work~\cite{Austin5} to a dynamical SMA model.

Figs.~\ref{qQuench}(a) and \ref{qQuench}(b) show that time traces of the phase $\theta$ provide a rigorous characterization of the dynamical phase diagram and DPTs, while in contrast, evolution of spin populations $\rho_{m_F}$ alone cannot directly identify what regime of the phase diagram the system is in without comparison to the theoretical phase diagram. We therefore focus on the temporal behavior of phase-based observables when studying DPTs in this work. 

For sharper identification of the DPTs in real time, we plot the extracted $\cos(\theta/2)$ in Fig.~\ref{qQuench}(c) for the two experiments shown in Figs.~\ref{qQuench}(a) and \ref{qQuench}(b) as well as a control set in which $q$ remains constant. For the two sets where $q$ was quenched, the experimental $\cos(\theta/2)$ begins significantly changing with time shortly after the quench, while in contrast $\cos(\theta/2)$ remains close to one for the control set (see Fig.~\ref{qQuench}(c)). These observations are consistent with SMA predictions: in the interaction regime $\cos(\theta/2)$ has a tiny peak-to-peak amplitude, while in the Zeeman regime $\cos(\theta/2)$ oscillates between $\pm1$. Therefore, a significant change in $\cos(\theta/2)$ indicates a DPT from the interaction to the Zeeman regime has occurred in our experiments. To quantify this behavior, we define the cut-off time $t_c$ as the time at which $\theta$ first satisfies $\abs{\theta}>\pi/2$, corresponding to $\cos(\theta/2)<0.7$, and becomes inconsistent with observations in the interaction regime (see Fig.~\ref{qQuench}(c)).

Strictly speaking, the rigorous detection of a DPT requires observing a nonanalytic change in an order parameter as a control parameter is varied. However, the similar dependence of $t_c$ and a typical order parameter $\beta$ on the control parameter $q/c_2$ confirms that $t_c$ can also be used to observe the DPT (see Fig.~\ref{tcVsBeta}(a)). Here $\beta=2-A_{\mathrm{pp}}$ with $A_{\mathrm{pp}}$ being the peak-to-peak amplitude of $\cos(\theta/2)$. The time needed to meaningfully measure $t_c$ and $\beta$, denoted $T_{t_c}$ and $T_{\beta}$ respectively, depends on the state during the quench. Notably, $T_{t_c}$ is predicted to be always shorter than or equal to $T_{\beta}$, as shown for a typical initial state in Fig.~\ref{tcVsBeta}(b).  Intuitively, $t_c$ can be determined from a handful of observations after the quench. In contrast, multiple observations over at minimum a full period of the spin dynamics are required to determine  commonly-used order parameters, e.g., $\beta$, or similar observables based on the time average or oscillation amplitude of $\cos(\theta/2)$~\cite{Austin5,Qingze2025,Guan2021}, winding number of $\theta$~\cite{Feldmann2021}, or the time average or steady-state behavior of $\rho_0$~\cite{Dag2018,Yingmei2019,Zhou2023,Qingze2025}. 

 \begin{figure*}[t]
	\includegraphics[width=176mm]{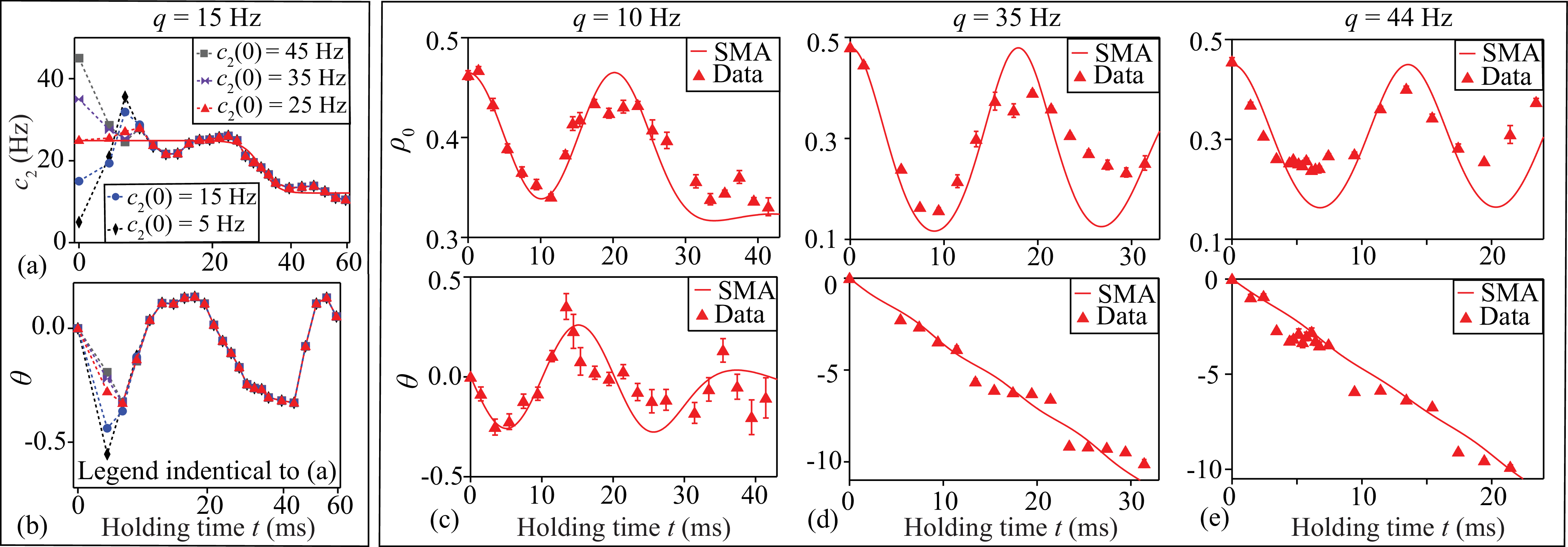}
    \caption{Markers display (a) $c_2(t)$ and (b) $\theta(t)$, extracted from observed spin population dynamics at $q=15~\mathrm{Hz}$, quickly converges despite drastically different initial guesses for $c_2(0)$. The red solid line in panel (a) is a sigmoidal fitting curve to the $c_2(0)=25~\mathrm{Hz}$ dataset. (c) Upper: triangles show the observed $\rho_0$ time evolution at $q=10$~Hz after the moving-lattice sequence. Lower: triangles display the corresponding $\theta$ extracted from the spin population dynamics  utilizing $c_2(t)$ based on the fitting shown in panel~(a). Lines in both subpanels display SMA predictions (see Eq.~\eqref{MF_Ham}) based on the fitting shown in panel~(a). (d, e) Similar to (c) but at (d) $q=35$~Hz and (e) $q=44$~Hz. }\label{DifferentQ}
\end{figure*}

\begin{figure}[tb]
		\includegraphics[width=86mm]{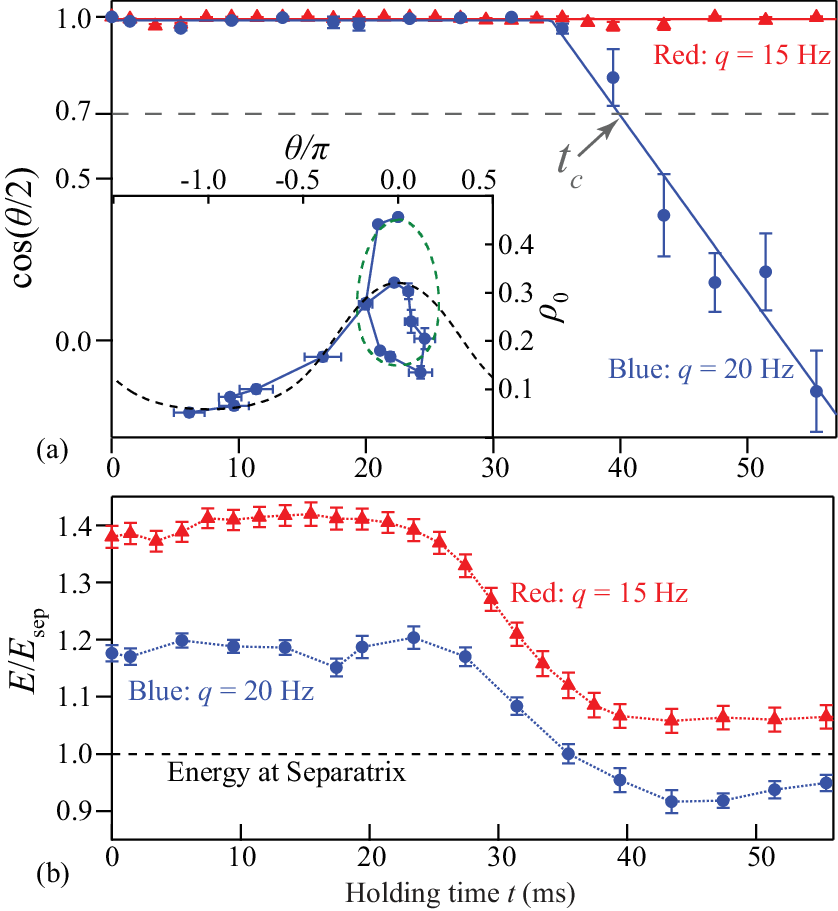}
		\caption{ (a) The appearance of a finite $t_c\approx40~\mathrm{ms}$ in the $q=20~\mathrm{Hz}$ dataset (circles) identifies a DPT driven by the moving-lattice-induced change in $c_2(t)$, while no DPT occurs in the $q=15~\mathrm{Hz}$ set  (triangles) as $\cos(\theta/2)>0.7$ for all $t$. Solid lines are linear fits while the dashed line marks $\cos(\theta/2)=0.7$. Inset: Circles display the DPT and the phase diagram for the $q=20~\mathrm{Hz}$ set shown in the main panel. The green (black) line displays the approximate equal energy contours for the interaction (Zeeman) regime occurring before (after) the drastic moving-lattice-induced change in $c_2$. (b) Blue circles (red triangles) show the extracted system energy $E$ drops below (stays above) the separatrix energy [dashed line] at $q=20~\mathrm{Hz}$ ($q=15~\mathrm{Hz}$) due to the moving-lattice-induced change in $c_2(t)$.
        } \label{DPT}
	\end{figure}     
\section{Spin-spatial-coupling induced DPTs}%
Having established that $\cos(\theta/2)$ distinguishes distinct dynamical regimes of $F=1$ spinor gases and $t_c$ more sharply identifies DPTs in real time for the case where the system is well understood in theory (see Fig.~\ref{qQuench}), we extend the analysis to a much more complicated moving-lattice system (see Fig.~\ref{DifferentQ}). This demonstrates the techniques applicability even with limited knowledge of system parameters or theory predictions. The moving-lattice sequence transfers a significant fraction of the atoms to the $\mathbf{p}=2\hbar \mathbf{k}_L$ momentum state, and these atoms are slowed as they move away from the minima of the ODT potential. The interplay between the ODT and the moving lattice causes complex spatial dynamics and rapid number loss, modifying the atomic density and the interactions governing spin dynamics~\cite{Zach1}.  

These violent spatial dynamics make standard methods of obtaining the value of $c_2$, whether from the estimated experimental atomic density or theoretical modeling, less reliable; as the spatial dynamics are highly sensitive to a large number of system parameters~\cite{Zach1}. Instead the observed spin dynamics can be utilized to reliably estimate $c_2$ through iteratively first solving Eq.~\eqref{RhoEqOfMotion} for $\theta$ and then Eq.~\eqref{ThetaEqOfMotion} for $c_2$, provided $\rho_0(t)$ is not close to 0.5 and the system remains in the interaction regime (see Fig.~\ref{DifferentQ}(a)). This method requires a time point where all relevant system parameters are known, and, while the initial state and $q$ can be precisely measured, $c_2(0)$ must be estimated. Figures~\ref{DifferentQ}(a) and \ref{DifferentQ}(b) demonstrate that the simultaneously extracted $c_2$ and $\theta$ curves rapidly converge regardless of the initial estimate for $c_2(0)$, confirming that the extracted values faithfully reflect the information carried in the spin population dynamics. The robustness of the extraction enables its use even in highly-nonequilibrium systems where system parameters may not be precisely known, for example in the moving-lattice system (see Figs.~\ref{DifferentQ}(c-e)).  

To extend the analysis across a broad range of $q$ in the moving-lattice system, we use an estimate of $c_2$, provided by the fitting curve (the solid line in Fig.~\ref{DifferentQ}(a)), to extract $\theta$ and predict the behavior of the system at other $q$. This transfer of the extracted interaction $c_2$ to other $q$ is supported by the similarities in the extracted $c_2(t)$ for data sets taken at different $q$ and also by similarities in the spatial
dynamics and the rate of atom number loss observed at all
q studied. Additional support is provided by theoretical simulations which show excellent agreement between the SMA predictions that utilize a time-dependent $c_2$ derived from scalar Gross-Pitaevskii (GP) simulations (in which $c_2$ is explicitly independent of $q$ and all spin states share a common time-dependent spatial mode) and the full spinor GP simulations (in which $c_2$ could potentially be $q$ dependent and multi-modal effects are allowed)~\cite{Zach1}. Although these simulations only qualitatively capture our experimental data because the exact time-dependence of $c_2$ is highly sensitive to numerous system parameters, the excellent agreement between the two simulations indicates that the fundamental assumptions of the SMA predictions, including that $c_2(t)$ is $q$ independent, are valid. We demonstrate this extension in Fig.~\ref{DifferentQ}(c) (\ref{DifferentQ}(d) and \ref{DifferentQ}(e)) which shows a typical example at $q=10~\mathrm{Hz}$ ($q=35~\mathrm{Hz}$ and $q=44~\mathrm{Hz}$), where the system remains in the interaction (Zeeman) regime throughout the dynamics with bounded (unbounded) $\theta$, and the dynamics are fairly well captured by the SMA predictions informed by the $c_2(t)$ fitting curve.

While datasets featured in Fig.~\ref{DifferentQ} remain in a single dynamical regime throughout the dynamics, a comparison of moving-lattice datasets taken at different $q$ (see Fig.~\ref{DPT}(a)) demonstrate that at an appropriate $q$ the system undergoes a DPT driven by the moving-lattice-induced change in $c_2$ from $c_2 \approx 25~\mathrm{Hz}$ at $t=0$ to $c_2\approx12~\mathrm{Hz}$ at $t\gg0$. As seen in Fig.~\ref{DPT}(a), after a long initial plateau near one the observable $\cos(\theta/2)$ drops below 0.7 for the $q=20~\mathrm{Hz}$ dataset (blue circles) at $t_c\approx40~\mathrm{ms}$ indicating that a DPT has occurred, while in contrast $\cos(\theta/2)$ remains close to one for all $t$ in the $q=15~\mathrm{Hz}$ dataset (red triangles) indicating no DPT occurs. The moving-lattice-tuned phase diagram and DPT at $q=20~\mathrm{Hz}$ can also be well described by equal energy contours predicted by Eq.~\eqref{MF_Ham} for a constant $c_2=25~\mathrm{Hz}\approx c_2(t=0)$ before (green dotted line) and $c_2=12~\mathrm{Hz}\approx c_2(t\gg0)$ after (black dotted line) an effective $c_2$ quench induced by the moving lattices (see Fig.~\ref{DPT}(a) inset). 

We can confirm these observations, i.e., that the moving-lattice system undergoes (does not undergo) a DPT when $q=20~\mathrm{Hz}$ $(q=15~\mathrm{Hz})$, by examining the energy $E$ of the system. Here $E$ can be readily evaluated from Eq.~\eqref{MF_Ham} once $\theta$ and $c_2$ are known. The separatrix that separates the Zeeman and interaction regimes in the phase diagram of $F=1$ spinor gases lies along the energy contour where $E=E_{\rm sep}$ and $E_{\rm sep}=h\cdot q$ if the magnetization $M=0$~\cite{Yingmei2009_2}. The energy $E$ provides a complementary view of the physics underlying the DPT, while $\cos(\theta/2)$ offers visually distinct behavior between the regimes without required knowledge of a potentially time-dependent $q$ such as in the data presented in Fig.~\ref{qQuench}. In Fig.~\ref{DPT}(b), both the $q=15~\mathrm{Hz}$ (red) and $q=20~\mathrm{Hz}$ (blue) data sets start firmly in the interaction regime where $E>E_{\rm sep}$. As $c_2$ decreases due to the violent spatial dynamics and atom number loss in the moving-lattice system, $E$ decreases correspondingly. For the $q=20~\mathrm{Hz}$ data set, this decrease in $E$ results in the system crossing the separatrix and undergoing a DPT, i.e., $E<E_{\rm sep}$ for $t\gtrsim35~\mathrm{ms}$; while for the $q=15~\mathrm{Hz}$ data set, $E$ remains larger than $E_{\rm sep}$ for all holding times studied and therefore the system does not undergo a DPT (see Fig.~\ref{DPT}(b)). Therefore the observations in Fig.~\ref{DPT} confirm that $\theta$, $t_c$, and $E$ can be used to detect DPTs in real time in a system with \textit{a priori} unknown system parameters utilizing the presented techniques.

\section{Discussion \& Outlook}
Our results demonstrate the real-time observation of DPTs in spinor gases using the system energy and phase-based observables extracted from spin population dynamics
both in free space and in a complex moving-lattice
system subject to unknown time-dependent interactions. The direct study of the temporal phase behavior as the system undergoes a DPT may have applications in understanding crossover phenomena and universality, with potential extensions to nonintegrable models. Additionally, our work introduces the cutoff time $t_c$ as an observable that can quickly identify DPTs at holding times when commonly-used order parameters still exhibit transient, nonuniversal behavior. We also demonstrate a robust method to extract time-dependent interaction values from spin population dynamics in highly-nonequilibrium systems. This method allows a better characterization of microscopic and effective Hamiltonian parameters, advancing the quantum simulation capabilities of spinor gases. The success of predictions and models based on these extracted interactions in explaining the complicated moving-lattice spin dynamics suggests that similar methods can be extended to other complex systems with time-dependent parameters, such as Floquet systems under a periodically driven magnetic field, driven interactions, or resonant spin-flopping fields.

    \begin{acknowledgments}
		\noindent{We acknowledge support from the Noble Foundation and the National Science Foundation through Grants No. PHY-2513302 and No. DGE-2510202. %
        TB acknowledges support by the Air Force Office of Scientific Research under Award No. FA9550-25-1-0340. 
       % in case we want to acknowledge NRT
       % NRT NSF Grant No. DGE-2510202. 
        }
	\end{acknowledgments}

\end{document}